\documentclass[twocolumn,showpacs,preprintnumbers,amsmath,amssymb,prb,superscriptaddress,10pt,aps]{revtex4-1}

\usepackage{natbib}
\usepackage{graphicx}
\usepackage{textcomp}
\usepackage{units}
\usepackage{dcolumn}
\usepackage{amsmath,amssymb}
\usepackage{verbatim} 
\usepackage[usenames,dvipsnames]{color}
\usepackage{bm} 

\newcommand{\superscript}[1]{\ensuremath{^{\textrm{#1}}}}

\begin{document}

\title{Detecting Ultrasound Vibrations by Graphene Resonators}

\author{G.J. Verbiest}
\affiliation{JARA-FIT and 2nd Institute of Physics, RWTH Aachen University, 52056 Aachen, Germany, EU}

\author{J.N. Kirchhof}
\affiliation{JARA-FIT and 2nd Institute of Physics, RWTH Aachen University, 52056 Aachen, Germany, EU}
\affiliation{Department of Physics, Freie Universit{\"a}t Berlin, 14195 Berlin, Germany, EU}

\author{J. Sonntag}
\affiliation{JARA-FIT and 2nd Institute of Physics, RWTH Aachen University, 52056 Aachen, Germany, EU}
\affiliation{Peter Gr{\"u}nberg Institute (PGI-8/9), Forschungszentrum J{\"u}lich, 52425 J{\"u}lich, Germany, EU}

\author{M. Goldsche}
\affiliation{JARA-FIT and 2nd Institute of Physics, RWTH Aachen University, 52056 Aachen, Germany, EU}
\affiliation{Peter Gr{\"u}nberg Institute (PGI-8/9), Forschungszentrum J{\"u}lich, 52425 J{\"u}lich, Germany, EU}

\author{T. Khodkov}
\affiliation{JARA-FIT and 2nd Institute of Physics, RWTH Aachen University, 52056 Aachen, Germany, EU}
\affiliation{Peter Gr{\"u}nberg Institute (PGI-8/9), Forschungszentrum J{\"u}lich, 52425 J{\"u}lich, Germany, EU}

\author{C. Stampfer}
\affiliation{JARA-FIT and 2nd Institute of Physics, RWTH Aachen University, 52056 Aachen, Germany, EU}
\affiliation{Peter Gr{\"u}nberg Institute (PGI-8/9), Forschungszentrum J{\"u}lich, 52425 J{\"u}lich, Germany, EU}
\email{stampfer@physik.rwth-aachen.de}

\begin{abstract}
Ultrasound detection
is one of the most important nondestructive subsurface characterization tools of materials, whose goal is to laterally resolve the subsurface structure with nanometer or even atomic resolution.
In recent years, graphene resonators attracted attention as loudspeaker and ultrasound radio, showing its potential to realize communication systems with air-carried ultrasound.
Here we show a graphene resonator that detects ultrasound vibrations propagating through the substrate on which it was fabricated.
We achieve ultimately a resolution of $\approx7$~pm/$\mathrm{\sqrt Hz}$ in ultrasound amplitude at frequencies up to 100~MHz.
Thanks to an extremely high nonlinearity in the mechanical restoring force, the resonance frequency itself can also be used for ultrasound detection.
We observe a shift of 120~kHz at a resonance frequency of 65~MHz for an induced vibration amplitude of 100~pm with a resolution of 25~pm.
Remarkably, the nonlinearity also explains the generally observed asymmetry in the resonance frequency tuning of the resonator when pulled upon with an electrostatic gate.
This work puts forward a sensor design that fits onto an atomic force microscope cantilever and therefore promises direct ultrasound detection at the nanoscale for nondestructive subsurface characterization.
\ \\ \ \\
Keywords: Graphene, ultrasound detection, NEMS, resonator, scanning probe microscopy
\ \\
\end{abstract}

\maketitle

The discovery of graphene \cite{novoselov2004electric} gave access to a new class of resonators that led to ultra-sensitive mass \cite{chaste2012}, charge \cite{lassagne2009,steele2009}, motion \cite{schmid2014}, and force sensors \cite{moser2013}, due to their high stiffness \cite{Lee385}, low mass density \cite{chen2013grapheneresonator,chen2009performance}, and low dissipation at low temperatures \cite{eichler2011}.
The unique electromechanical coupling in graphene allows for elegant electrical read-out of its resonator properties \cite{chen2009performance,sazonova2004tunableresonator,song2012}, which is preferred for the integration into electronic circuits.
This in combination with a typical resonance frequency in the order of tens of MHz makes graphene an interesting material for ultrasound microphones and radios \cite{zhou2013,zhou2015,woo2017}, and thus for ultrasound communication systems with air-carried ultrasound.
However, for applications such as nondestructive material testing and medical imaging \cite{hedrick2005ultrasound,szabo2004diagnostic}, the ultrasound propagates through a solid medium and is detected at a surface.
The ultrasound detectors in these applications suffer from the diffraction limit \cite{szabo2004diagnostic,castellini2009laser}, which prevents their application down to the nanoscale.
Yet, it is the nanoscale that gets increasingly important due to the ever ongoing miniaturization of electrical and mechanical components.
This problem was partly overcome by introducing ultrasound detection into an atomic force microscope (AFM) \cite{kolosov1993,yamanaka1996,cuberes2000,garcia2012}, which allowed the nondestructive visualization of buried nanostructures \cite{hu2011,vitry2015,kimura2013,shekhawat2005,cantrell2007,tetard2010,garcia2010,verbiest2016}.
As the resonance frequency of AFM cantilevers is much smaller than the typical MHz frequency of the ultrasound, the ultrasound detection relies on large ultrasound amplitudes ($\sim 0.1$ to $\sim 1$ nm) and the nonlinear interaction between the cantilever-tip and the sample to obtain a signal at a measurable frequency.
This makes quantitative measurements tedious as a detailed understanding of the nonlinear interaction \cite{sarioglu2004,parlak2008,rabe1996}, the resonance frequencies of the cantilever \cite{verbiest2016b}, and the indirect ultrasound pick up \cite{verbiest2015,bosse2014,verbiest2013,verbiest2013b,forchheimer2012} is required.
Consequently, nondestructive material testing and medical imaging with an AFM would greatly benefit from a direct (linear) detection scheme for the ultrasound vibration of the sample.
Graphene resonators seem ideal candidates, as their typical dimensions allow for the integration into standard AFM cantilevers and they have been shown to detect air-carried ultrasound.

\begin{figure}[!t]
    \centering
    \includegraphics[trim=0 0 0 0, clip, width=80mm]{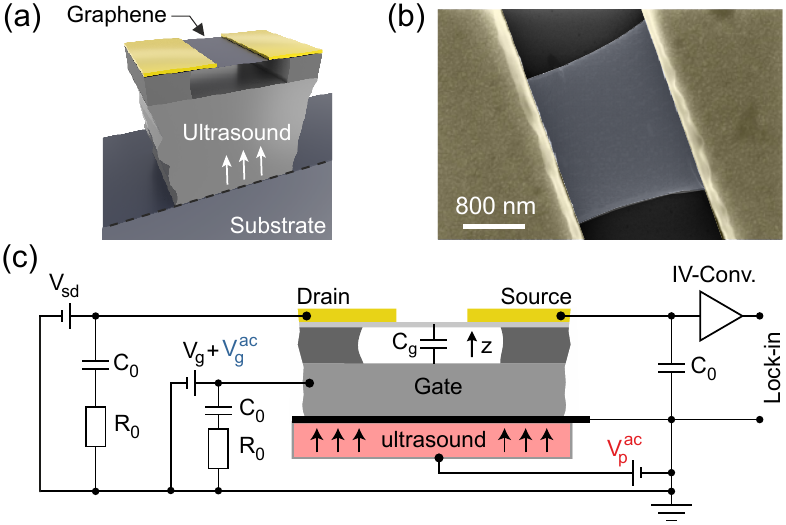}
    \caption{
    (a) Illustration of a graphene-based ultrasound detector and (b) a false-color scanning electron microscope image of a fabricated graphene resonator.
    The ultrasound propagates through the substrate to the gold contacts (dark yellow) and actuates the graphene resonator (light gray) which generates an electrical signal.
    (c) The detection circuit measures the resonance frequencies of the graphene resonator by the amplitude modulated down-mixing technique (see text).
    The resistances $R_0 = 50$ $\Omega$ and capacitances $C_0 = 100$ nF are for impedance matching and reducing the noise on the input of the IV-converter.}
    \label{fig1}
\end{figure}

Here we show graphene resonators that detect ultrasound vibrations introduced through the substrate.
We compare the response of the graphene resonators to ultrasound in a purely mechanical actuation scheme with that of the standard capacitive actuation (Figure~\ref{fig1}).
The vibration amplitude is measured in both actuation schemes with an amplitude modulated down-mixing technique \cite{chen2009performance}.
We show that both the vibration amplitude away from the resonance frequency of the graphene and the resonance frequency itself can be used for ultrasound detection with a resolution of 20-25~pm.
Interestingly, the nonlinearity underlying the ultrasound detection at the resonance frequency also explains the generally observed asymmetric tuning of it when pulling on the resonator with an electrostatic gate.

%
We fabricated graphene resonators on a standard $\mathrm{Si/SiO_2(285~nm)}$ substrate by mechanically exfoliating graphene and using electron beam lithography to pattern the electrical contacts.
The contacts consist of 5~nm Cr and 120~nm Au and also serve as clamps for the graphene resonator.
Finally, the graphene membrane is suspended by etching the $\mathrm{SiO_2}$ away in a 1\% hydrofluoric acid solution followed by a critical point dryer step to prevent the graphene membrane from collapsing due to capillary forces.
Using this method, we fabricated in total five devices (see Supporting Discussion~1). Devices D1-D4 consist of suspended graphene membranes while device D5 is based on a suspended hexagonal boron nitride (hBN)/graphene heterostructure.
In the following, we mainly focus on device~D1.
From the scanning electron microscope image (SEM) in Figure~\ref{fig1}b, we extract a width $W$ of 2.2~\textmu m and a distance $L$ between the clamps of 1.6~\textmu m.
Further optical, SEM, and AFM characterization is found in Figure~S1a-c.
The $\mathrm{SiO_2}$ is 140~nm underneath the graphene membrane, which results in a parallel plate capacitance $C_\text{g}$ for the back gate of 0.18~fF.
Note that the contacts are under etched over the same distance.
The capacitance $C_\text{g}$ allows us to extract (a lower bound of) the room temperature carrier mobility of 1.500~$\mathrm{cm^2/Vs}$ for holes and 1.000~$\mathrm{cm^2/Vs}$ for electrons from the conductance $G$ and transconductance $\partial G/\partial V_\text{g}$ tuning with gate potential $V_\text{g}$ (see Figure~S1d-e).
Raman spectroscopy measurements showed no D-peak and a narrow 2D-peak ($\sim 25$ cm$^\text{-1}$), hence indicating good quality of the graphene resonator (see Figure~S1f-g).
The average pre-strain in this resonator is 0.24\% (see Figure~S1h).

\begin{figure*}[!t]
    \centering
	\includegraphics[trim=0 0 0 0, clip, width=170mm]{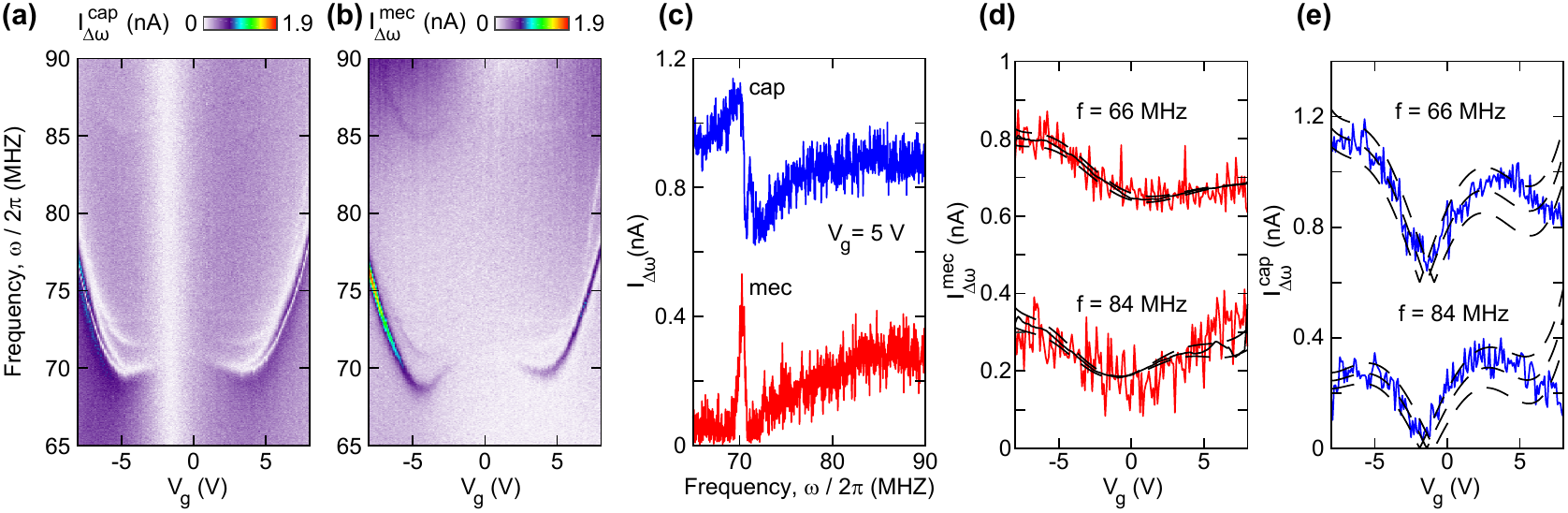}
    \caption{
    Unprocessed down-mixing current $I_{\Delta\omega}$ as a function of actuation frequency and applied gate potential $V_\text{g}$ in case of (a) the standard capacitive actuation $I_{\Delta\omega}^\text{cap}$ and (b) the purely mechanical actuation scheme $I_{\Delta\omega}^\text{mec}$. The line traces in (c) show $I_{\Delta\omega}$ at a gate potential of $V_\text{g} = 5$ V for the standard capacitive actuation (blue) and the purely mechanical actuation (red) scheme. Note that the position of the resonance frequency of the graphene resonator is approximately the same in both actuation schemes. By analyzing $I_{\Delta\omega}$ as a function of $V_\text{g}$ far away from the resonance frequency of the graphene resonator, we observe a background current. In the purely mechanical actuation scheme (d), this current allows us to quantify the ultrasound vibration of the contacts (see text). This contribution is completely covered by a background current in the standard capacitive actuation scheme (e). The black lines in (d) and (e) correspond to the fit of $I_{\Delta\omega}$ to Eq.~\ref{eq2} including the confidence interval due to the uncertainty in the transconductance (see Figure S1).
    }
	\label{fig2}
\end{figure*}

\begin{figure*}[!t]
    \centering
	\includegraphics[trim=0 0 0 0, clip, width=170mm]{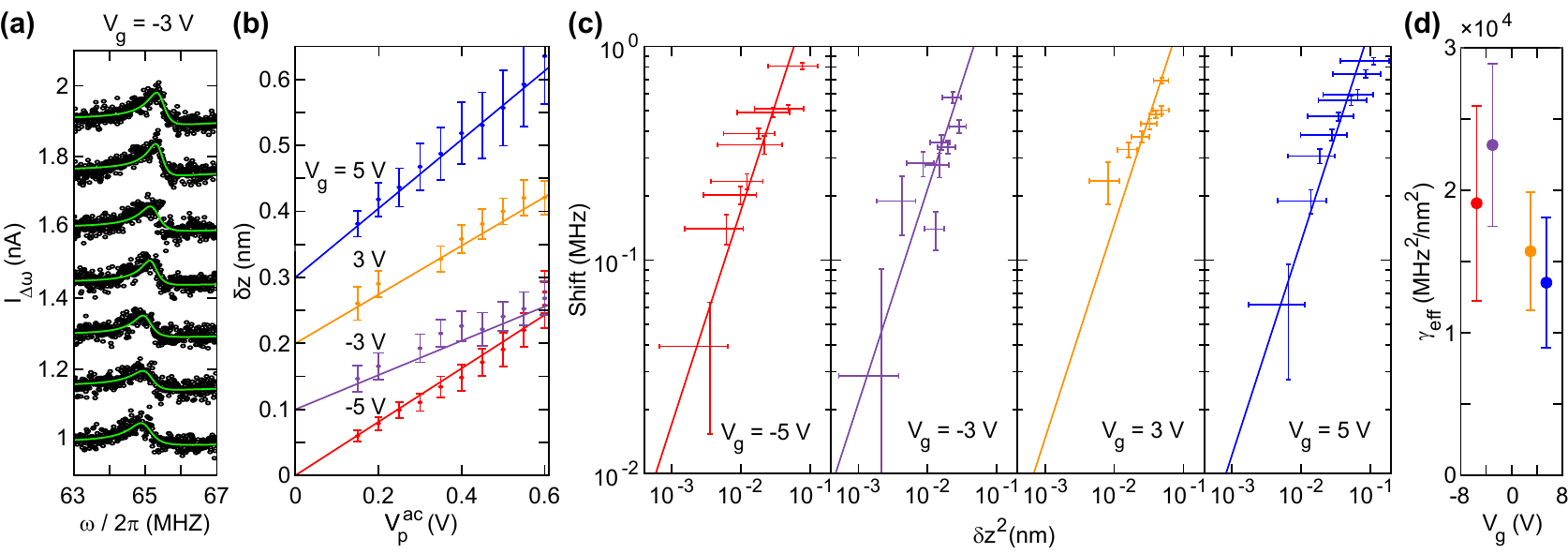}
    \caption{(a) $I_{\Delta\omega}$ for increasing drive amplitudes from bottom to top with the corresponding fit to a nonzero phase Lorentzian (green). The curves are offset with 0.15~nA for clarity. (b) The vibration amplitude $\delta z_c$ as a function of the drive voltage $V_\text{p}^\text{ac}$ on the piezoelectric element used for the mechanical actuation. The curves are offset by 0.1 nm for clarity. (c) The resonance frequency linearly shifts to higher frequencies scaling with the vibration amplitude squared. The slopes of these curves define the nonlinearity $\gamma_\text{eff}$ (d). Within error bars, the nonlinearity does not depend on the applied $V_\text{g}$.}
	\label{fig3}
\end{figure*}

In addition to the standard capacitive actuation of the resonator with a potential $V_\text{g}^\text{ac}$ on the gate, we use a piezoelectric element with a resonance frequency of 4.5~MHz (Figure~S2 and Ref.~\onlinecite{verbiest2015b}) to mechanically shake the substrate and thus introduce ultrasound to the graphene resonator.
The piezoelectric element is electrically isolated from the gate and is actuated with a voltage $V_\text{p}^\text{ac}$.
The actuation frequency $\omega/2\pi$ is swept around the mechanical resonance frequency of the suspended graphene membrane.
In both cases, a small modulation potential $V_\text{sd} = 10$~mV at frequency $\omega/2\pi \pm \Delta\omega/2\pi$ is applied across the graphene membrane to generate a current $I_{\Delta\omega}$ at frequency $\Delta\omega/2\pi$, which is amplified with an IV-converter and measured with an UHF lock-in amplifier from Z\"{u}rich Instruments (Figure~\ref{fig1}c).
We tune $V_\text{g}^\text{ac}$ and $V_\text{p}^\text{ac}$ such that we have approximately the same $I_{\Delta\omega}$ in both actuation schemes.
Note that all experiments were performed in a vacuum of $10^{-5}$~mbar at room temperature.

Figures~\ref{fig2}a and \ref{fig2}b show the measured current $I_{\Delta\omega}$ as a function of $\omega/2\pi$ and $V_\text{g}$ for the standard capacitive actuation and the purely mechanical actuation.
We observe in both actuation schemes the resonance frequencies as dips and peaks in $I_{\Delta\omega}$ at approximately the same positions (compare the two traces in Figure~\ref{fig2}c).


The current $I_{\Delta\omega}$ is described by \cite{chen2009performance}:
\begin{equation}
I_{\Delta\omega} = V_{ds}  \frac{\partial G}{\partial V_\text{g}} \left( V_\text{g}^\text{ac} + \left( V_\text{g} - V_\text{cnp} \right) \frac{\partial_z C_\text{g}}{C_\text{g}} \delta z \right),\label{eq2}
\end{equation}
in which $\delta z$ is the displacement of the graphene membrane that also includes the time-dependent displacement of the contacts, and $V_\text{cnp}$ is the charge neutrality potential of the graphene membrane.
The capacitance $C_\text{g}$ and $\partial_z C_\text{g}$ are given by the zeroth and the first order term in $\delta z$ of a series expansion of the parallel plate approximation such that $C_\text{g}/\partial_z C_\text{g} = 175$~nm (for $\delta z = 0$).
Note that $I_{\Delta\omega}$ should be zero when $\partial G/\partial V_\text{g}$ is zero.
The fact that $I_{\Delta\omega}$ does not truly become zero (Figure~\ref{fig2}a-b) indicates the presence of a very small electrical cross-talk in our setup.
This generates an offset current, which is independent of $V_\text{g}$ and is subtracted before fitting $I_{\Delta\omega}$ with Eq.~\ref{eq2}.
Far away from the mechanical resonance of the graphene membrane, $\delta z$ reduces to the vibration amplitude $\delta z_c$ of the contacts.
To extract $\delta z_c$, we assume that $\delta z_c$ does not depend on $V_\text{g}$~[\onlinecite{chen2009performance}].
This leaves $\delta z = \delta z_c$ and an effective $V_\text{g}^\text{ac}$ as fitting parameters in Eq.~\ref{eq2} as $\partial G/\partial V_\text{g}$ and $V_\text{cnp}$ are measured independently and the potentials $V_\text{g}$ and $V_\text{ds}$ are fixed in the experiment.
The current $I_{\Delta\omega}$ is thus split into an even part in $(V_\text{g}-V_\text{cnp})$ that is proportional to $(\partial G/\partial V_\text{g})(V_\text{g}-V_\text{cnp})$ and an odd part that is proportional to $(\partial G/\partial V_\text{g})$.
The former is characterized by $\delta z = \delta z_c$ and the latter by an effective $V_\text{g}^\text{ac}$.

In case of the purely mechanical actuation far away from the resonance frequencies of the graphene, we find a $\delta z_c$ of 30 and 65~pm at 66~MHz and 84~MHz (Figure~\ref{fig2}d).
The effective $V_\text{g}^\text{ac}$ is well below 0.5\% of the set $V_\text{p}^\text{ac}$ even when $V_\text{p}^\text{ac} = 1.5$~V (see Table~S1).
This confirms the good electrical isolation of the piezoelectric element from the gate.
Note that the phase between the offset current and $\delta z = \delta z_c$ shifts by 180 degrees when increasing $\omega/2\pi$ from below to above the mechanical resonance frequency of the graphene membrane.
Consequently, $I_{\Delta\omega}$ is larger at 66~MHz than at 75~MHz for $V_\text{g} = -5$~V whereas this is opposite at $V_\text{g} = 5$~V.
The apparent even behavior of $I_{\Delta\omega}$ in $(V_\text{g}-V_\text{cnp})$ (Figure~\ref{fig2}d) confirms the dominantly mechanical origin of $I_{\Delta\omega}$.
The functional form of $I_{\Delta\omega}$ remains the same over the full measured frequency range, even when crossing a mechanical resonance of the graphene (see Figure~S3).
In addition, the drive amplitudes of the mechanical resonance frequencies of the graphene are one order of magnitude too small to account for the observed $\delta z_c$ (see below).
This suggests that factors such as the clamping, membrane size, and ultrasound wavelength are important for unravelling the exact relation between the measured $\delta z_c$ and the impinging one, i.e. vibration amplitude of the piezoelectric element.
For devices D2-D5 (see Figures S4-S7 for data similar to Figure~\ref{fig2}), we found a similar response to the ultrasound as the one presented here.
The graphene resonators thus detect ultrasound vibrations in a frequency range that covers at least two orders of magnitude: from 1~MHz to 100~MHz.
In case of the capacitive actuation, the nonzero $V_\text{g}^\text{ac}$ results in a large background current that covers the signal coming from the vibrating contacts.
This results in an upper bound for the vibration amplitude $\delta z = \delta z_c$ of $\sim 40$~pm at 66 and 84~MHz (Figure~\ref{fig2}e).

Let us next consider the vibration amplitude at the resonance frequency of the graphene membrane.
If the graphene membrane is pulled down by the static gate potential $V_\text{g}$, the inversion symmetry is broken.
The tension in the graphene gets reduced when moving the membrane away from the gate, whereas the tension is increased when the membrane is moved towards the gate.
This symmetry breaking is described by the nonlinear restoring force $m \beta z^2$.
In addition to this term, we take into account the well-known Duffing nonlinearity $m \gamma z^3$.
Here, $\beta$ and $\gamma$ are constants quantifying the strength of the symmetry breaking effect and the Duffing nonlinearity in the equation of motion ($m\ddot{z}^2 + m\Gamma\dot{z} + kz + m \beta z^2 + m \gamma z^3 = F_\text{d}$, where $\Gamma$ is the linewidth quantifying the damping and $F_\text{d}$ is the effective driving force).
The nonlinear terms containing $\beta$ and $\gamma$ lead to a vibration amplitude dependent frequency shift \cite{landau2004}:
\begin{equation}
\Delta \omega_0 = \frac{3}{8} \frac{\gamma_{\mathrm{eff}}}{\omega_0} \delta z^2,
\label{eq3}
\end{equation}
\noindent
in which $\Delta \omega_0/2\pi$ is the shift in resonance frequency $\omega_0/2\pi$ and $\gamma_{\mathrm{eff}} = \gamma - 10\beta^2/(9\omega_0^2)$.

To quantify $\gamma_{\mathrm{eff}}$, we measured the resonance frequency as a function of vibration amplitude $\delta z$ at a fixed gate potential $V_\text{g}$ by varying the ultrasound drive potential $V_\text{p}^\text{ac}$ (Figure~\ref{fig3}a-b).
The resonance frequency is extracted by fitting the current $I_{\Delta\omega}$ with a nonzero phase Lorentzian:
\begin{equation}
I_{\Delta\omega} = \frac{A \omega_0^2 \cos{\left(\frac{\omega^2-\omega_0^2}{\omega\omega_0/Q}+\theta\right)}}{\sqrt{\left(\omega^2-\omega_0^2\right)^2 +\left(\omega\omega_0/Q\right)^2}},
\label{eq5}
\end{equation}
in which $A = F_\text{d}/m$ is the effective drive amplitude, $\theta$ is the non-zero phase, and $Q = \omega_0/\Gamma$ its quality factor.
The vibration amplitude at resonance is given by the current $A Q$, which we translate into $\delta z$ using the transconductance and the measurement parameters in Eq.~\ref{eq2}.

The linear dependence of the resonance frequency $\omega_0/2\pi$ with $\delta z^2$ shown in Figure~\ref{fig3}c is in agreement with Eq.~\ref{eq3}.
We find that $\gamma_{\mathrm{eff}} = 18.000 \pm 5.000~\mathrm{MHz^2/nm^{2}}$ (Figure~\ref{fig3}d).
The nonlinearity $\gamma_{\mathrm{eff}}$ is positive and thus the resonance frequency increases with increasing vibration amplitude.
We conclude that the Duffing nonlinearity $\gamma$ dominates over $\beta$.
According to literature, this is consistent with the relatively large pre-strain ($\sim 0.24\%$) in the graphene resonator \cite{eichler2011,eichler2013}.

\begin{figure*}[!t]
    \centering
	\includegraphics[trim=0 0 0 0, clip, width=170mm]{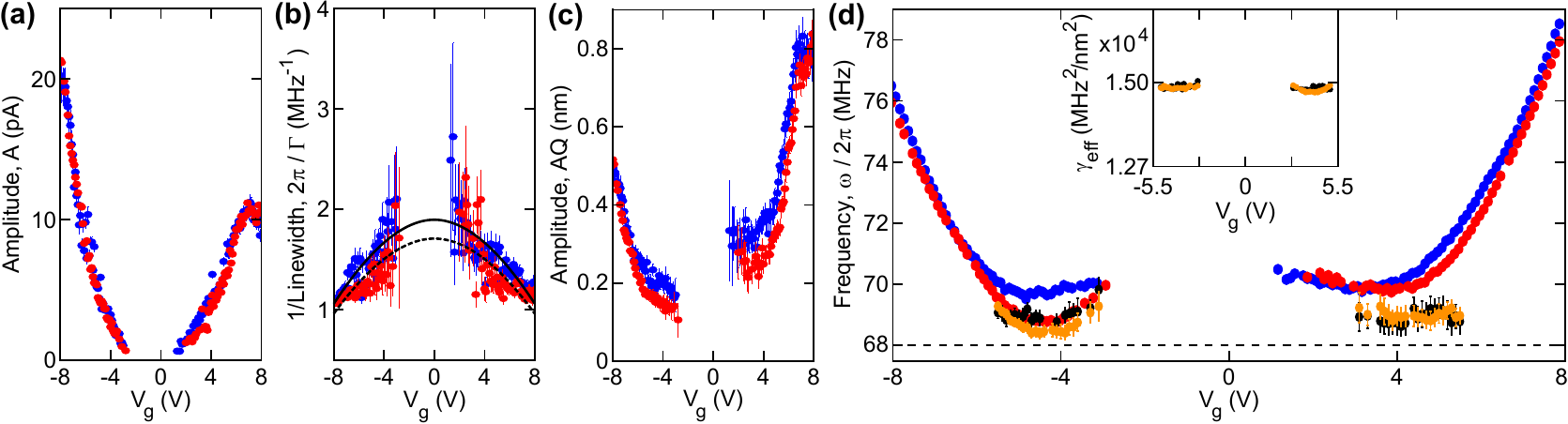}
    \caption{The extracted (a) effective drive amplitude $A$, (b) inverse linewidth $2\pi/\Gamma$, and (c) vibration amplitude $A Q$ at the resonance frequency. The black lines in (b) indicate that the inverse linewidth is in agreement with Joule heating (see text). In all panels, the standard capacitive actuation is depicted in blue and the purely mechanical actuation in red. Note that the vibration amplitude of the membrane in nanometer is significantly lower in the purely mechanical actuation case. (d) The resonance frequency in the standard capacitive actuation (blue) appears more symmetric than the one extracted from the purely mechanical actuation (red). This is a consequence of the vibration amplitude (c) and allows for an alternative way of determining the nonlinearity $\gamma_\text{eff}$. As the electrostatic force dictates a symmetric resonance frequency in $V_\text{g}$, we can subtract the effect of a finite vibration amplitude from the measured resonance frequencies: standard capacitive actuation (black) and purely mechanical actuation (orange). The inset depicts the estimated $\gamma_\text{eff}~=~14.800$ MHz\superscript{2}/nm\superscript{2} with an uncertainty of $~\pm~2.300$ MHz\superscript{2}/nm\superscript{2} illustrated by the size of the box.}
	\label{fig4}
\end{figure*}

The effective drive amplitude $A$ determined from the $I_{\Delta\omega}$ maps in Figure~\ref{fig2}a-b is approximately the same for both actuation schemes (Figure~\ref{fig4}a).
Surprisingly, the linewidth $\Gamma$ depicted in Figure~\ref{fig4}b is significantly lower in the case of mechanical actuation.
The tuning of the inverse linewidth $2\pi/\Gamma$ with $V_\text{g}$ is in both actuation schemes well captured by Joule heating \cite{song2012}:
\begin{equation}
\frac{2\pi}{\Gamma}=\frac{2\pi}{\Gamma_0}-\frac{\alpha V_\text{g}^2}{1+\eta\left|V_\text{g}-V_\text{cnp}\right| },
\label{eq6}
\end{equation}
in which $\eta$ parameterizes the change in conductance of the graphene with applied $V_\text{g}$ and $V_\text{cnp} = -1.3$~V (Figure~S1d), $\alpha$ describes the change in inverse linewidth, and $\Gamma_0$ is the linewidth at $V_\text{g} = 0$. We extract $\eta = 0.025$~V\superscript{-1} from the conductance in Figure~S1. The fit to Eq.~\ref{eq6} gives for both actuation schemes an $\alpha$ of $0.014\pm 0.001$~s/V\superscript{2}. In contrast, there is a significant difference in $\Gamma_0$: the capacitive actuation scheme gives $\Gamma_0/2\pi = 528 \pm 56$~kHz whereas $\Gamma_0/2\pi = 587 \pm 61$~kHz in the mechanical actuation scheme.
We attribute this difference to the larger contact vibration amplitude in the purely mechanical actuation scheme (see above), as predicted by theoretical work \cite{kim2009,jiang2012}.

Due to the difference in quality factors $Q$, the amplitude at the resonance frequency significantly differs.
We find that the graphene membrane vibrates with amplitudes up to 0.8~nm (Figure~\ref{fig4}c).
Interestingly, the vibration amplitude is not symmetric in $|V_\text{g}|$.
This is due to the transconductance $\partial G/\partial V_\text{g}$, which is symmetric in $|V_\text{g}-V_\text{cnp}|$ (Figure S1), and the quality factor, which is roughly symmetric in $|V_\text{g}|$.
In combination with the nonlinearity $\gamma_\text{eff}$, this leads to an apparent asymmetric dependence of the extracted resonance frequency in $|V_\text{g}|$ (Figure~\ref{fig4}d), whereas the electrostatic force dictates a symmetric behavior as it depends on $V_\text{g}^2$.

The measured resonance frequencies in combination with the measured vibration amplitudes gives us yet another way of determining the nonlinearity $\gamma_\text{eff}$.
We extract the nonlinearity $\gamma_\text{eff}$ by demanding a symmetric behavior of the resonance frequencies in $V_\text{g}$ after subtracting the effect of the vibration amplitude, which results in $\gamma_\text{eff}~=~14.800~\pm~2.300$~MHz\superscript{2}/nm\superscript{2} (inset~Figure~\ref{fig4}d).
This value for $\gamma_\text{eff}$ is in agreement with the $\gamma_{\mathrm{eff}} = 18.000 \pm 5.000~\mathrm{MHz^2/nm^{2}}$ extracted from the measurement in which the drive amplitude was varied (Figure~\ref{fig3}).
Therefore, we conclude that the nonlinearity $\gamma_\text{eff}$ in combination with the vibration amplitude of the graphene resonator explains the generally observed asymmetry in the resonance frequency tuning with $V_\text{g}$ \cite{chaste2012,steele2009,moser2013,chen2013grapheneresonator,chen2009performance,eichler2011,eichler2013,sazonova2004tunableresonator,song2012,song2012}.

We can also make use of the nonlinearity $\gamma_\text{eff}$ for ultrasound detection.
The nonlinearity is so strong that even a small vibration amplitude of 100~pm shifts the resonance frequency by 120~kHz.
This suggest that after calibrating the nonlinearity, one can measure the ultrasound impinging on the resonator by monitoring its resonance frequency.
In our measurements, we have an average measurement error of 29~kHz on the extracted resonance frequency, which translates into a detection resolution in ultrasound amplitude of 25~pm, making this method promising for ultrasound detection.

\begin{figure}[!t]
    \centering
	\includegraphics[trim=0 0 0 0, clip, width=79.6mm]{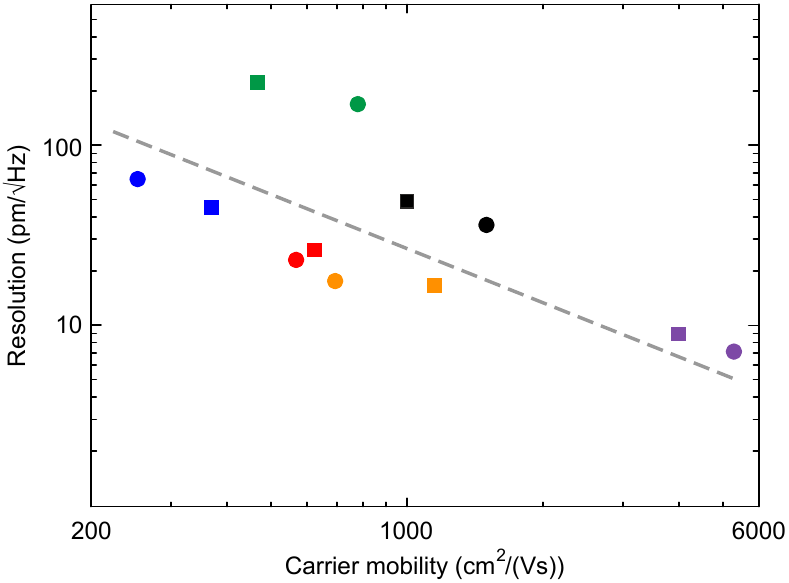}
    \caption{
    The experimentally extracted resolution in detecting the ultrasound vibration as a function of the (lower limit of the) carrier mobility. Each color represents a different sample. Device D1 presented in the manuscript is shown in black. Data from three more graphene devices are shown in green (D2), red (D3), and purple (D4). The data shown in blue and orange was obtained on a hBN/graphene heterostructure (D5) before and after current annealing. Figures S4-S7 provide raw data of devices D2-D5. The circles (squares) circles represent the resolution at $V_\text{g}-V_\text{cnp} = 5(-5)$ V. The dashed gray line has a slope of -1 and indicates that the resolution is inversely proportional to the mobility.}
	\label{fig5}
\end{figure}

To gain insight into the achievable resolution, we examine the detection sensitivity $S_{\delta z}^{-1} = V_{ds}  \tfrac{\partial G}{\partial V_\text{g}} \left(V_\text{g} - V_\text{cnp}\right) \tfrac{\partial_z C_\text{g}}{C_\text{g}}$ to measure $\delta z$.
Note that $S_{\delta z}$ is completely determined by the electrical properties of the graphene sheet and its distance to the gate as well as the experimentally set $V_{ds}$ and $V_\text{g}$.
This quantity is thus {\it independent} of the mechanical response of the graphene sheet.
Consequently, the detection sensitivity is optimized by maximizing $\tfrac{\partial G}{\partial V_\text{g}}$ and $V_\text{g} - V_\text{cnp}$.
Therefore, the resolution, which is $S_{\delta z}$ multiplied by the noise in $I_{\Delta\omega}\approx50-120$~pA (see Figure~\ref{fig1}c-e), should be best for the sample with the highest mobility.
Figure~\ref{fig5} summarizes the experimentally extracted resolution as a function of mobility for all measured devices.
For each device, we extracted the resolution at $V_\text{g} - V_\text{cnp} = \pm5$~V.
The dashed gray line indicates that the resolution is inversely proportional to the mobility.

In summary, we investigated the feasibility of detecting ultrasound with a graphene resonator.
We introduced a purely mechanical actuation scheme to measure the ultrasound vibration amplitude of the electrical contacts of the graphene.
This new actuation scheme is also applicable to other two-dimensional resonators and even provides a method for actuating nonconductive resonators.
Using the presented devices, we are able to sense the ultrasound vibration amplitude of the electrical contacts up to at least 100~MHz with a resolution of $\approx7$~pm/$\mathrm{\sqrt Hz}$.
Alternatively, we can use the mechanical nonlinearity $\gamma_\text{eff}~=~14.800~\pm~2.300$ MHz\superscript{2}/nm\superscript{2} of the graphene resonator to detect ultrasound by measuring the resonance frequency.
Due to the nonlinearity, we can detect ultrasound via resonance frequency monitoring with a resolution of 25~pm.
We showed that graphene resonators can directly pick-up ultrasound at frequencies inaccessible for near field probes.
Thus, this work presents a first step to ultrasound detection at the nanoscale using graphene.
Although the best sensitivity we have shown in this work (60~pm/nA) is inferior to the sensitivity of an STM (10~pm/nA) \cite{chen2007STM}, there is room for further improvement in sample quality. For example, if the carrier mobility is improved by a factor of 10 (which is plausible when encapsulating graphene in hBN \cite{Bolotin2008351,luca2016}), our scheme could fully compete with STM. When integrating a graphene-based resonator device on the back of an AFM cantilever this would open the door high-sensitive ultrasound detection at the nanoscale on arbitrary substrates and surfaces.

\section*{Associated content}
\subsection*{Supporting Information}
\noindent
Details on the sample characterization and on the ultrasound transducer are available free of charge via the Internet at \url{http://pucs.acs.org}.

\section*{Author information}
\subsection*{Corresponding author}
\noindent
E-mail: stampfer@physik.rwth-aachen.de


\subsection*{Notes}
\noindent
The authors declare no competing financial interests.

\section*{Acknowledgments}
\noindent
Support by the ERC (GA-Nr. 280140), the Helmholtz Nanoelectronic
Facility (HNF) \cite{hnf2017} at the Forschungszentrum J\"ulich, and the Deutsche Forschungsgemeinschaft (DFG) (SPP-1459) are gratefully acknowledged.
G.V. acknowledges funding by the Excellence Initiative of the German federal and state governments.

\providecommand{\latin}[1]{#1}
\makeatletter
\providecommand{\doi}
  {\begingroup\let\do\@makeother\dospecials
  \catcode`\{=1 \catcode`\}=2 \doi@aux}
\providecommand{\doi@aux}[1]{\endgroup\texttt{#1}}
\makeatother
\providecommand*\mcitethebibliography{\thebibliography}
\csname @ifundefined\endcsname{endmcitethebibliography}
  {\let\endmcitethebibliography\endthebibliography}{}

\newpage
\clearpage

\onecolumngrid
\setcounter{figure}{0}
\renewcommand{\figurename}{Figure}
\renewcommand{\tablename}{Table}
\renewcommand{\thetable}{S\arabic{table}}
\renewcommand{\thefigure}{S\arabic{figure}}

{\Large {\bf Supporting information:\\Detecting Ultrasound Vibrations by Graphene Resonators}}

\newpage

Supplementary Discussion 1: Additional samples
\ \\ \ \\
In addition to the device D1 shown in the main manuscript, we fabricated four more devices (D2-D5).
Three of these devices (D2-D4) consist of a suspended single layer graphene membrane and one (D5) of a suspended hBN/graphene heterostructure.
These devices have a width ranging from 2.9 to 3.1 \textrm{$\mu$}m and a length ranging from 1.3 to 1.6 \textrm{$\mu$}m.
We measured on all these devices the response of the resonator to capacitive actuation and purely mechanical actuation for frequencies between 1 MHz and 100 MHz (see Figures~S4-S7).
These measurements were performed with a higher $V_\text{mod}$ and $V_\text{g}^\text{AC}$ in comparison to the one presented in the main manuscript (see Table~S1) to maximize the down-mixing current $I_{\Delta\omega}$.
Consequently, the background currents obtained on these samples are higher than the one in Figure~2 of the main manuscript.

\ \\ \ \\ \ \\ \ \\ \ \\ \ \\ \ \\ \ \\
\ \\ \ \\ \ \\ \ \\ \ \\ \ \\ \ \\ \ \\
\ \\ \ \\ \ \\ \ \\ \ \\ \ \\ \ \\ \ \\

\newpage

\begin{figure}[th]
  \includegraphics[width=85mm]{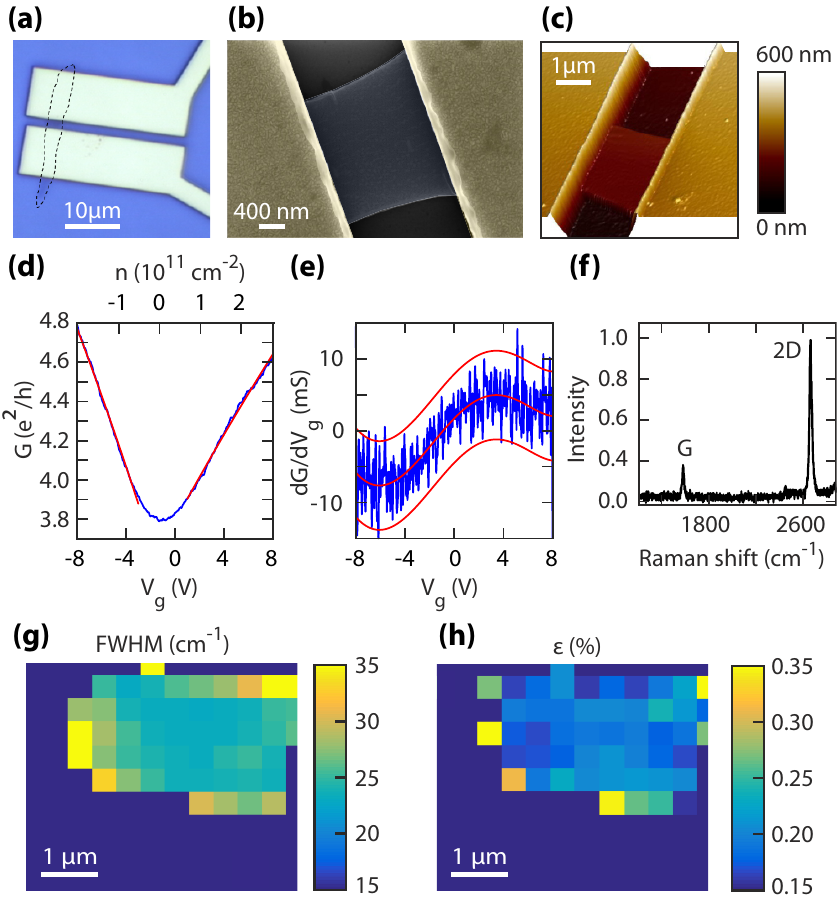}
  \caption{
    (a) optical, (b) scanning electron microscope, and (c) atomic force microscope image from which we extract the physical dimensions of the suspended area: $2.2\times1.6$ \textrm{$\mu$}m\superscript{2}, as well as the graphene-gate distance: 140 nm. (d) conductance and (e) transconductance as a function of applied gate voltage allows an estimation of the two-terminal hole mobility of 1.500 cm\superscript{2}/Vs and electron mobility of 1.000 cm\superscript{2}/Vs. (f) typical Raman spectra of the suspended area does not show a defect peak. (g) the narrow width of 2D-mode shows the good quality of the resonator. (h) the extracted pre-strain $\epsilon$ is on average 0.24\%.
  }
  \label{sup_fig1}
\end{figure}

\ \\ \ \\ \ \\ \ \\ \ \\ \ \\ \ \\ \ \\
\ \\ \ \\ \ \\ \ \\ \ \\ \ \\ \ \\ \ \\
\ \\ \ \\ \ \\ \ \\ \ \\ \ \\ \ \\ \ \\

\newpage

\begin{figure}[th]
\includegraphics[width=85mm]{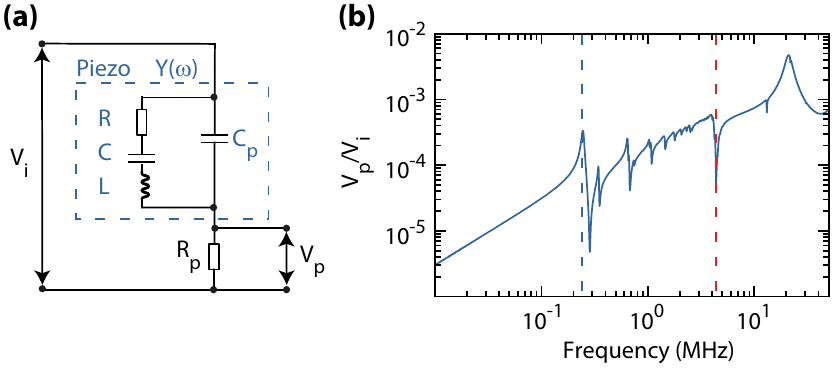}
  \caption{
    (a) Electric circuit used to determine the resonance frequencies of the piezoelectric element \cite{verbiest2015b}.
    The applied voltage $V_i$ is set to 1 V.
    We measured the voltage $V_p$ over a low Ohmic resistance of $R_p = 4.7$~$\Omega$, to determine the admittance $Y(\omega)$ as a function of the excitation frequency.
    The piezoelectric element is modeled as a capacitance $C_p$ in parallel to a RCL-circuit.
    Each mechanical resonance frequency of the piezoelectric elements is described by a corresponding RCL-circuit.
    (b) The measured transfer function $V_0/V_i$ of the piezoelectric element as a function of frequency.
    Each mechanical resonance is observed as a dip and peak in the otherwise linearly increasing transfer function $V_0/V_i$.
    The natural resonance frequencies $f_{xy}$ of the piezoelectric element can be calculated with $f_{xy} = N_{xy} n / L$, in which $N_{xy}$ is the frequency constant of the of a specific mode, L is the physical length of the piezoelectric element in direction of the motion, and $n = 1, 2, 3...$ is the mode number \cite{pic}.
    The blue dashed line marks the lowest order resonance of the longitudinal mode at around 130 kHz and the red dashed line indicates the transversal mode at 4.5MHz, which is relevant
    for the mechanical excitation of the resonator.
  }
  \label{sup_fig2}
\end{figure}

\ \\ \ \\ \ \\ \ \\ \ \\ \ \\ \ \\ \ \\
\ \\ \ \\ \ \\ \ \\ \ \\ \ \\ \ \\ \ \\
\ \\ \ \\ \ \\ \ \\ \ \\ \ \\ \ \\ \ \\

\newpage

\begin{figure}[!t]
    \centering
    \includegraphics[trim=0 0 0 0, clip, width=170mm]{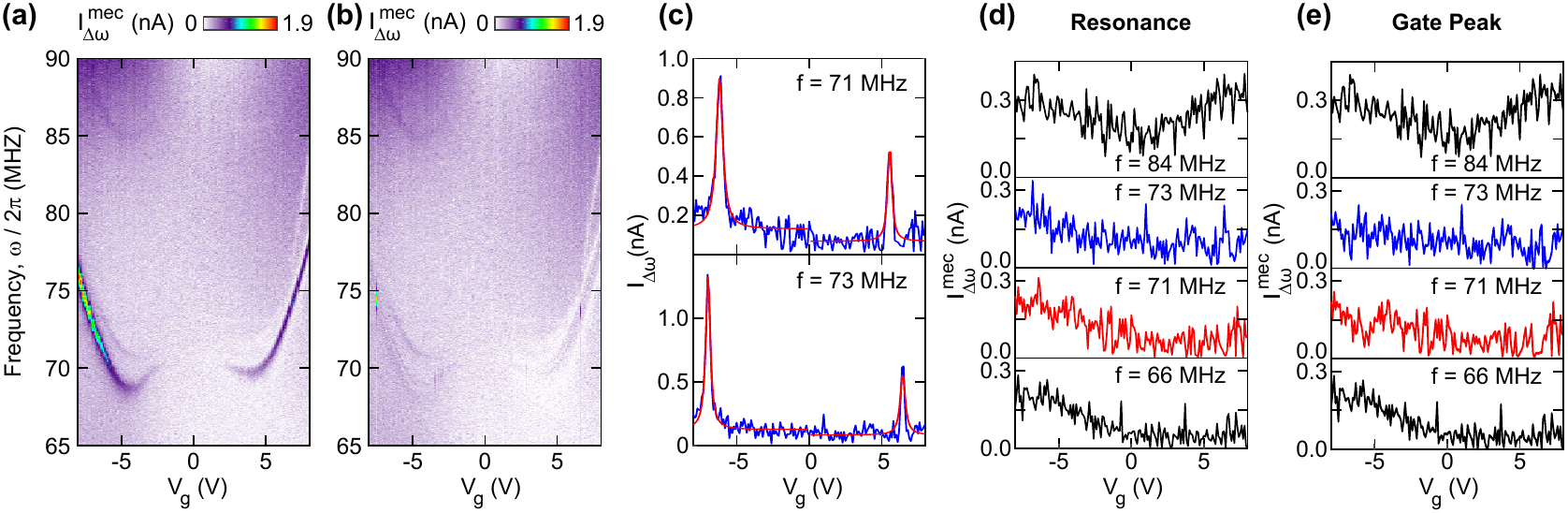}
    \caption{
    (a) Unprocessed mechanical down-mixing current $I_{\Delta\omega}^\text{mec}$ as a function of actuation frequency $f = \omega/2\pi$ and applied gate potential $V_\text{g}$ (see Figure~2(a) of the main manuscript). (b) $I_{\Delta\omega}^\text{mec}$ after subtraction of the first mechanical resonance frequency of the graphene by using the fit results in Figure~4 of the main manuscript. The subtraction shows that a substantial background current remains. (c) In a second approach, we extract a line trace (blue) at a fixed frequency from panel (a) and fit the peaks with a Lorentzian curve (red). Both approaches allows the unambiguous determination of the background current, i.e. the $I_{\Delta\omega}^\text{mec}$ not originating from the mechanical resonance of the graphene resonator. Panel (d) shows the background current for different frequencies obtained after subtraction of the first mechanical resonance of the graphene and (e) depicts the background current using the method outlined in panel (c). The unprocessed $I_{\Delta\omega}^\text{mec}$ at 66 and 84 MHz have been added for comparison (see Figure~2(d) of the main manuscript). The great similarity between panels (d) and (e) show that the functional form of $I_{\Delta\omega}$ remains the same over the full measured frequency range, even when crossing a mechanical resonance of the graphene.}
    \label{sup_fig3}
\end{figure}

\ \\ \ \\ \ \\ \ \\ \ \\ \ \\ \ \\ \ \\

\newpage

\begin{figure}[!t]
    \centering
    \includegraphics[trim=0 0 0 0, clip, width=170mm]{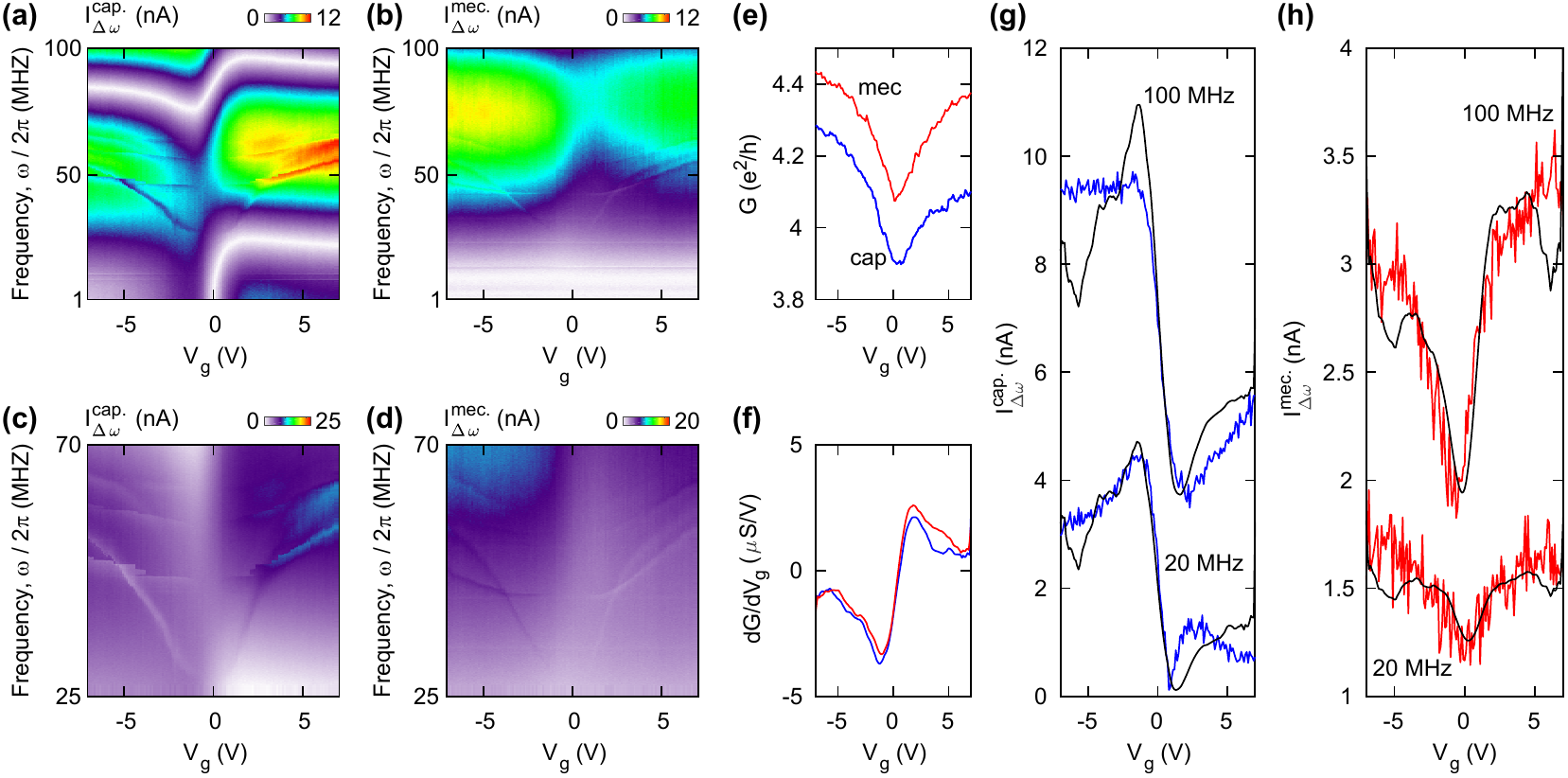}
    \caption{
    Device D2: single layer graphene membrane. Unprocessed down-mixing current $I_{\Delta\omega}$ as a function of actuation frequency and applied gate potential $V_\text{g}$ in case of (a) the standard capacitive actuation $I_{\Delta\omega}^\text{cap}$ and (b) the purely mechanical actuation scheme $I_{\Delta\omega}^\text{mec}$. The maps in (c) and (d) show a zoom in of $I_{\Delta\omega}$ for the capacitive and purely mechanical actuation to highlight the resonance frequencies. These panels illustrate the tuning with applied gate potential $V_\text{g}$ of the resonance frequencies and to compare the capacitive actuation result with that of the purely mechanical actuation. Panel (e) and (f) show the conductance $G$ and the transconductance $d$ for the capacitive (blue) and the purely mechanical actuation (red) scheme. By analyzing $I_{\Delta\omega}$ as a function of $V_\text{g}$ far away from the resonance frequency of the graphene resonator, we observe a background current.
    Panel (g) shows the background current at 20 and 100~MHz for the capacitive actuation and panel (h) for the purely mechanical actuation. The black lines in (g) and (h) correspond to the fit of $I_{\Delta\omega}$ with Eq.~1 in the main manuscript.}
    \label{sup_fig4}
\end{figure}

\ \\ \ \\ \ \\ \ \\ \ \\ \ \\ \ \\ \ \\

\newpage

\begin{figure}[!t]
    \centering
    \includegraphics[trim=0 0 0 0, clip, width=170mm]{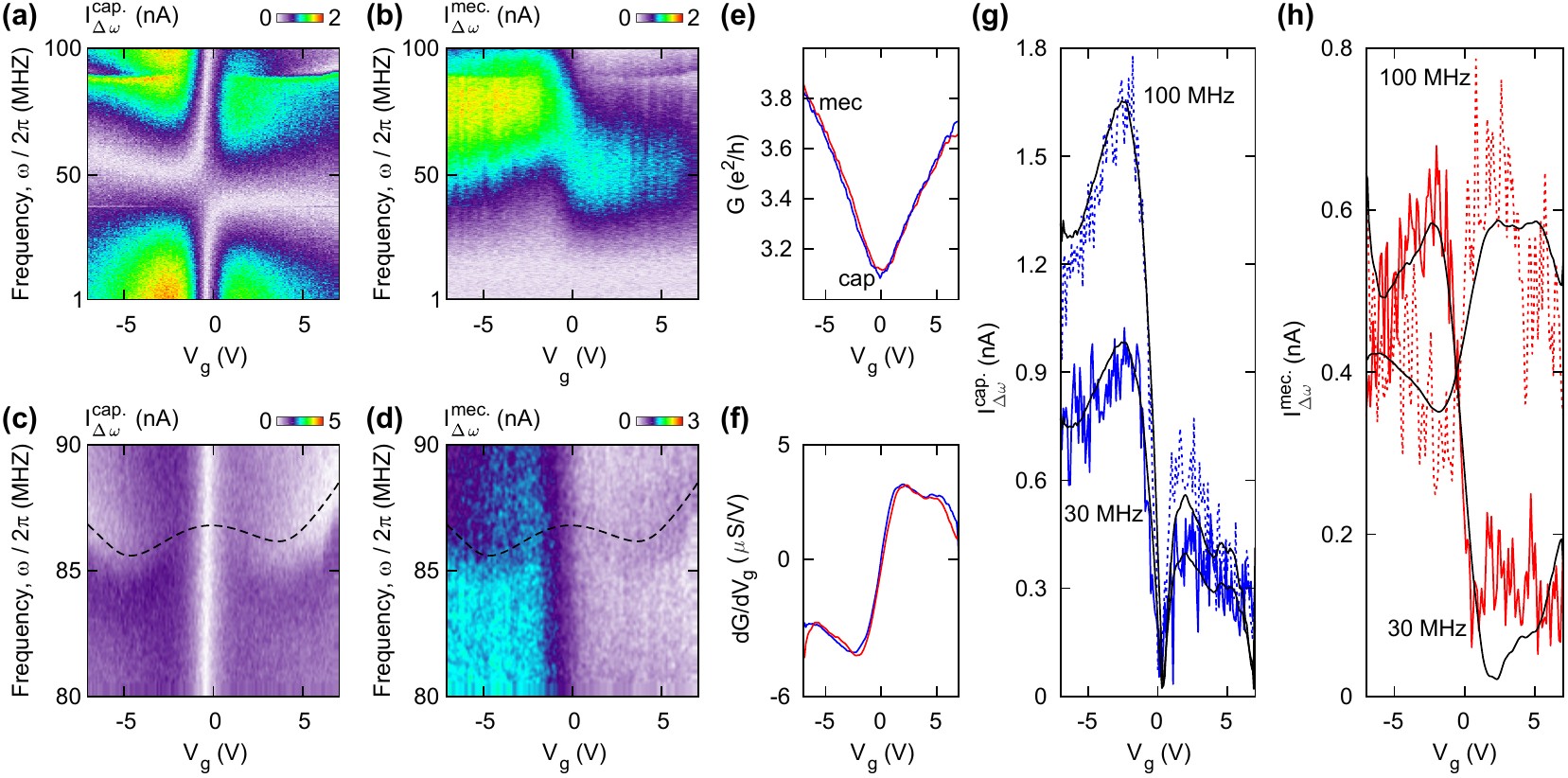}
    \caption{
    Device D3: single layer graphene membrane. Unprocessed down-mixing current $I_{\Delta\omega}$ as a function of actuation frequency and applied gate potential $V_\text{g}$ in case of (a) the standard capacitive actuation $I_{\Delta\omega}^\text{cap}$ and (b) the purely mechanical actuation scheme $I_{\Delta\omega}^\text{mec}$. The maps in (c) and (d) show a zoom in of $I_{\Delta\omega}$ for the capacitive and purely mechanical actuation to highlight the resonance frequencies. These panels illustrate the tuning with applied gate potential $V_\text{g}$ of the resonance frequencies and to compare the capacitive actuation result with that of the purely mechanical actuation. The black dashed line serves as a guide to the eye and is the same in both panels. Panel (e) and (f) show the conductance $G$ and the transconductance $d$ for the capacitive (blue) and the purely mechanical actuation (red) scheme. By analyzing $I_{\Delta\omega}$ as a function of $V_\text{g}$ far away from the resonance frequency of the graphene resonator, we observe a background current.
    Panel (g) shows the background current at 20 and 100~MHz for the capacitive actuation and panel (h) for the purely mechanical actuation. The black lines in (g) and (h) correspond to the fit of $I_{\Delta\omega}$ with Eq.~1 in the main manuscript.}
    \label{sup_fig5}
\end{figure}

\ \\ \ \\ \ \\ \ \\ \ \\ \ \\ \ \\ \ \\

\newpage

\begin{figure}[!t]
    \centering
    \includegraphics[trim=0 0 0 0, clip, width=170mm]{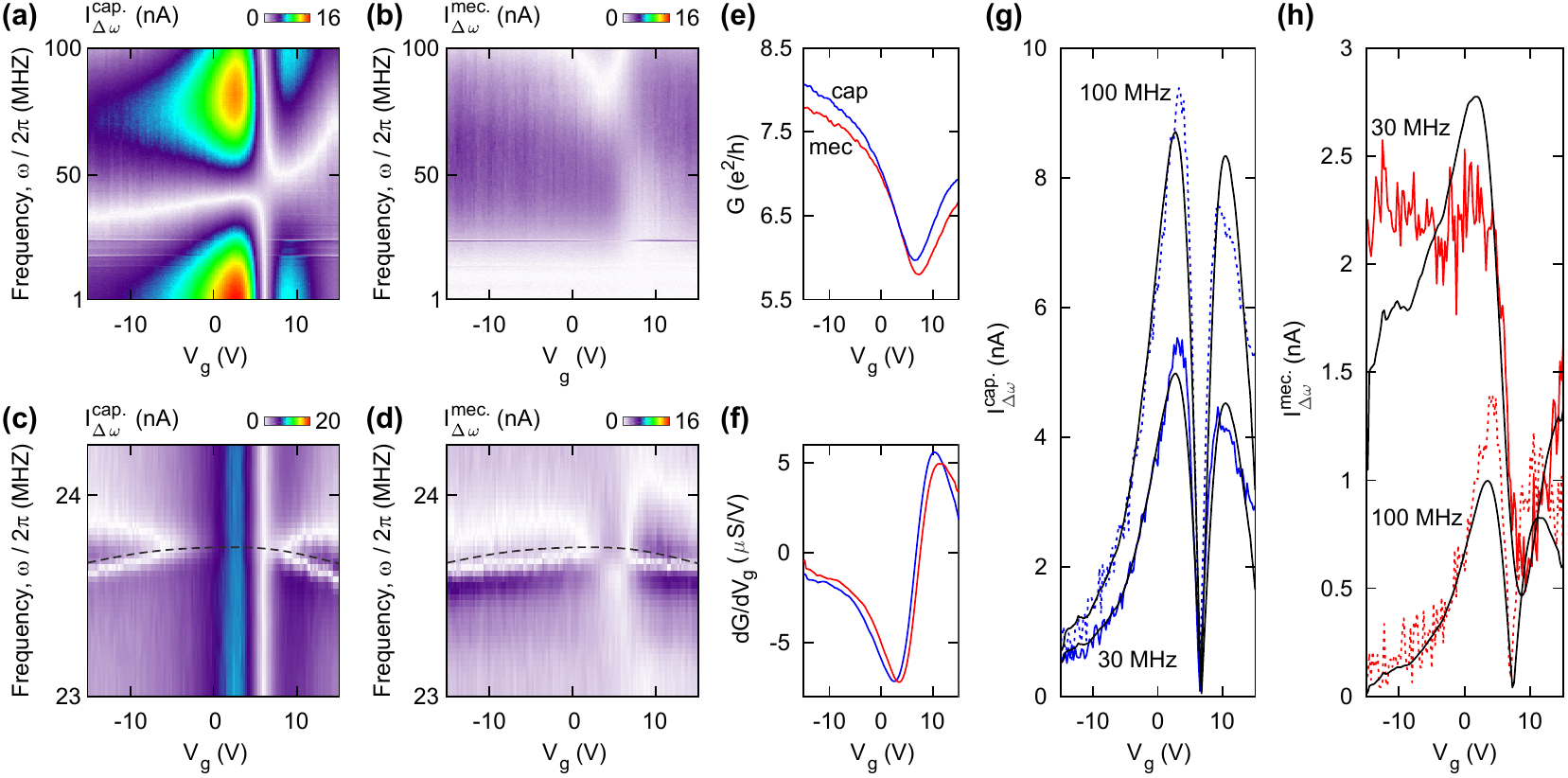}
    \caption{
    Device D4: single layer graphene membrane. Unprocessed down-mixing current $I_{\Delta\omega}$ as a function of actuation frequency and applied gate potential $V_\text{g}$ in case of (a) the standard capacitive actuation $I_{\Delta\omega}^\text{cap}$ and (b) the purely mechanical actuation scheme $I_{\Delta\omega}^\text{mec}$. The maps in (c) and (d) show a zoom in of $I_{\Delta\omega}$ for the capacitive and purely mechanical actuation to highlight the resonance frequencies. These panels illustrate the tuning with applied gate potential $V_\text{g}$ of the resonance frequencies and to compare the capacitive actuation result with that of the purely mechanical actuation. The black dashed line serves as a guide to the eye and is the same in both panels. Panel (e) and (f) show the conductance $G$ and the transconductance $d$ for the capacitive (blue) and the purely mechanical actuation (red) scheme. By analyzing $I_{\Delta\omega}$ as a function of $V_\text{g}$ far away from the resonance frequency of the graphene resonator, we observe a background current.
    Panel (g) shows the background current at 20 and 100~MHz for the capacitive actuation and panel (h) for the purely mechanical actuation. The black lines in (g) and (h) correspond to the fit of $I_{\Delta\omega}$ with Eq.~1 in the main manuscript.}
    \label{sup_fig6}
\end{figure}

\ \\ \ \\ \ \\ \ \\ \ \\ \ \\ \ \\ \ \\

\newpage

\begin{figure}[!t]
    \centering
    \includegraphics[trim=0 0 0 0, clip, width=170mm]{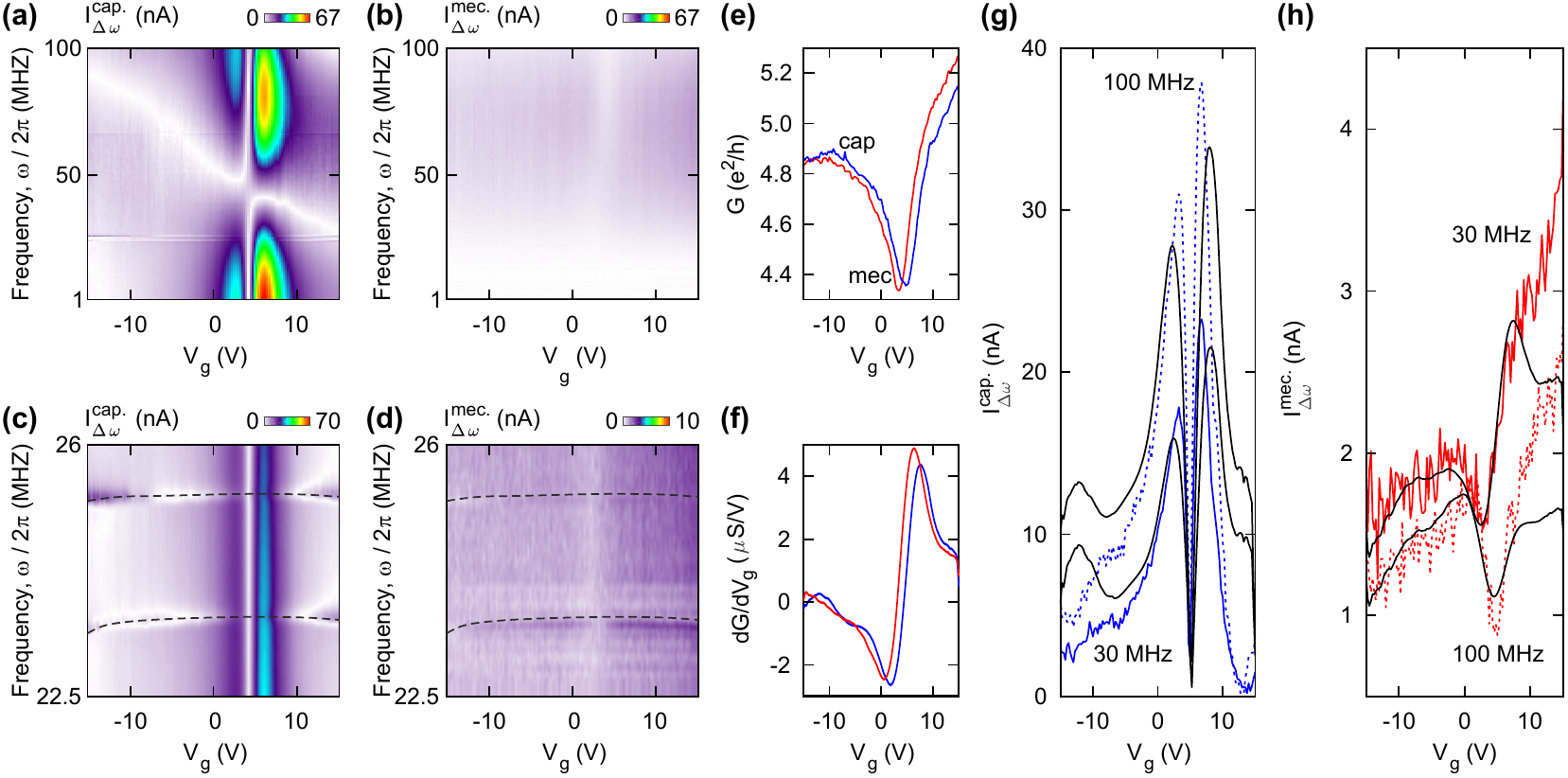}
    \caption{
    Device D5: graphene/hBN membrane (hBN thickness $\approx 5$~nm) after current annealing. Unprocessed down-mixing current $I_{\Delta\omega}$ as a function of actuation frequency and applied gate potential $V_\text{g}$ in case of (a) the standard capacitive actuation $I_{\Delta\omega}^\text{cap}$ and (b) the purely mechanical actuation scheme $I_{\Delta\omega}^\text{mec}$. The maps in (c) and (d) show a zoom in of $I_{\Delta\omega}$ for the capacitive and purely mechanical actuation to highlight the resonance frequencies. These panels illustrate the tuning with applied gate potential $V_\text{g}$ of the resonance frequencies and to compare the capacitive actuation result with that of the purely mechanical actuation. The black dashed line serves as a guide to the eye and is the same in both panels. Panel (e) and (f) show the conductance $G$ and the transconductance $d$ for the capacitive (blue) and the purely mechanical actuation (red) scheme. By analyzing $I_{\Delta\omega}$ as a function of $V_\text{g}$ far away from the resonance frequency of the graphene resonator, we observe a background current.
    Panel (g) shows the background current at 20 and 100~MHz for the capacitive actuation and panel (h) for the purely mechanical actuation. The black lines in (g) and (h) correspond to the fit of $I_{\Delta\omega}$ with Eq.~1 in the main manuscript.}
    \label{sup_fig7}
\end{figure}

\ \\ \ \\ \ \\ \ \\ \ \\ \ \\ \ \\ \ \\
\ \\ \ \\ \ \\ \ \\ \ \\ \ \\ \ \\ \ \\
\ \\ \ \\ \ \\ \ \\ \ \\ \ \\ \ \\ \ \\

\newpage

\renewcommand{\figurename}{Table}
\setcounter{figure}{0}
\begin{figure}[!t]
    \centering
    \includegraphics[trim=0 0 0 0, clip, width=80mm]{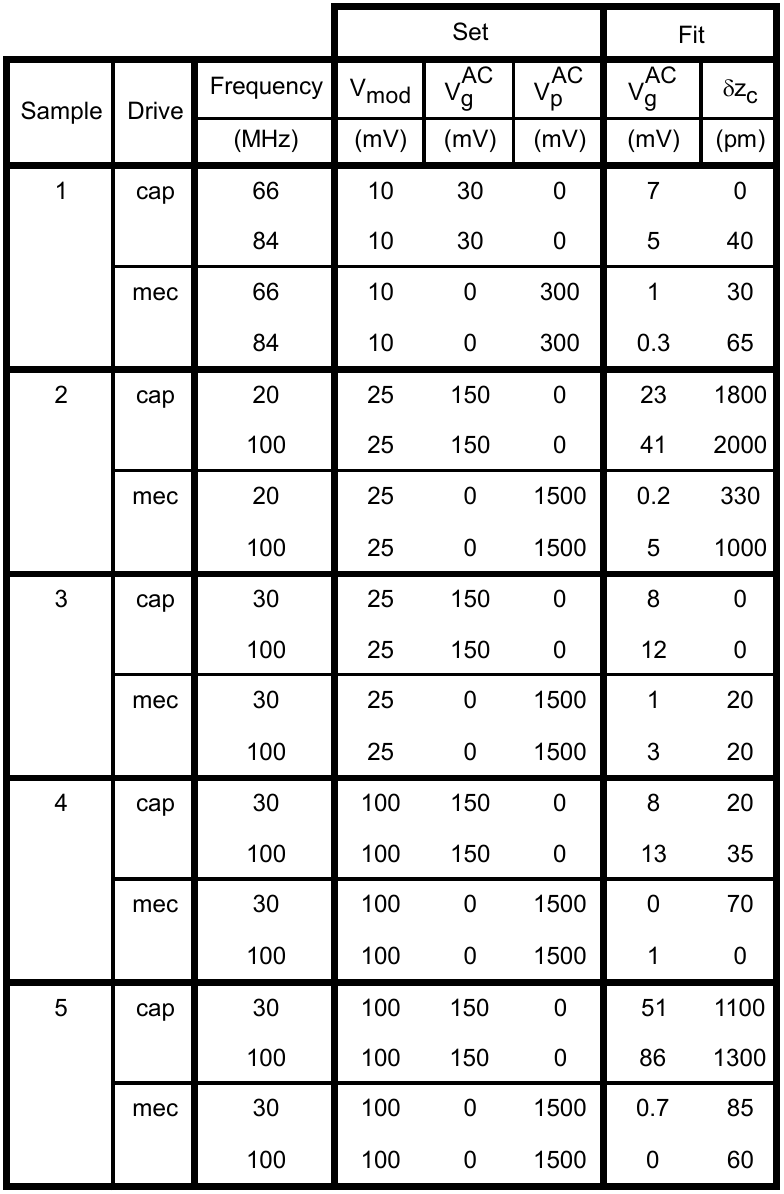}
    \caption{
    Table containing the fit parameters of the background currents depicted in Figure 2 of the main manuscript and in Figures S4-S7. The first column specifies the sample and the second one the type of actuation: capacitive (cap) or purely mechanical (mec). The third column lists the frequencies at which the background currents were recorded. The fourth to sixth column specify the experimentally set $V_\text{mod}$, $V_\text{g}^\text{AC}$, and $V_\text{p}^\text{AC}$. The last two columns denote the estimated $V_\text{g}^\text{AC}$ and $\delta z_\text{c}$. Note that for all purely mechanical actuation fits, the estimated $V_\text{g}^\text{AC}$ is well below 1\% of the set $V_\text{p}^\text{AC}$, which illustrates the good electrical isolation of the piezoelectric element from the gate.}
    \label{sup_tab1}
\end{figure}

\ \\ \ \\ \ \\ \ \\ \ \\ \ \\ \ \\ \ \\

\newpage

\providecommand{\latin}[1]{#1}
\makeatletter
\providecommand{\doi}
  {\begingroup\let\do\@makeother\dospecials
  \catcode`\{=1 \catcode`\}=2 \doi@aux}
\providecommand{\doi@aux}[1]{\endgroup\texttt{#1}}
\makeatother
\providecommand*\mcitethebibliography{\thebibliography}
\csname @ifundefined\endcsname{endmcitethebibliography}
  {\let\endmcitethebibliography\endthebibliography}{}

\end{document}